\documentstyle[11pt,newpasp,twoside,epsf]{article}
\def\edcomment#1{\iffalse\marginpar{\raggedright\sl#1\/}\else\relax\fi}
\reversemarginpar

\begin{document}
\title{Soft Cores in Late-Type Dwarf and LSB Galaxies from H$\alpha$ Observations}
 \author{D. Marchesini$^{1}$, E. D'Onghia$^{1,2}$, G. Chincarini$^{2,3}$,
C. Firmani$^{2}$, P.~Conconi$^{2}$, E. Molinari$^{2}$, A. Zacchei$^{4}$}
\affil{1-Universit\`a degli Studi di Milano, Via
Celoria 16, 20133, Milano, Italy;2-Osservatorio Astronomico di Brera-Merate,
Via Bianchi 46, 23807 Merate (LC), Italy;3-Universit\`a degli Studi di
Milano-Bicocca, Piazza dell'Ateneo Nuovo 1, 20126, Milano,
Italy;4-Telescopio Nazionale Galileo (TNG)}

\begin{abstract}
We present high spatial resolution H$\alpha$ rotation curves of late-type
dwarf and LSB galaxies. From our analysis we find good agreement between our
H$\alpha$ data and the
HI observations taken from the li\-te\-rature, concluding that the HI rotation curves for these
galaxies suffer very little from beam smearing. A preliminary analysis of our
data rules out the CDM model in the inner regions of these galaxies.
\end{abstract}

\section{Introduction}

The current scenario for structure formation in the universe is based on
the inflationary Cold Dark Matter (CDM) theory. High-resolution N-body
simulations predict, for the virialized haloes, density profiles with central cusps with $\rho \propto
r^{-1.5}$ when r $\to$ 0 (Moore et al. 1999), even more cuspy than previous
simulation results (Navarro, Frenk, and White 1997, hereafter NFW). These profiles are in conflict
with the observations: rotation curves of late-type dwarf and LSB ga\-laxies (DM
dominated galaxies) seem to rule out singular halo profiles and are in good agreement
with density profiles characterized by soft cores, central regions with
constant density (de Blok \& McGaugh 1997). On the other hand, van den Bosch \& Swaters (2001) claim that
these HI rotation curves are affected by beam smearing, concluding that the
observed rotation curves are consistent with NFW model. A good spatial
resolution (easily achievable with H$\alpha$ observations) is necessary to
explore the galaxy inner regions and rule out beam smearing. We show some
preliminary results of our H$\alpha$ observations for some dwarf and LSB
galaxies. This work assumes $\mathrm{H}_{o}$=75 km s$^{-1}$ Mpc$^{-1}$.

\section{Observations}

Observations of late-type dwarf and LSB galaxies were carried out at
the 3.6~m Telescopio Nazionale Galileo (TNG) both in
photometric and spectroscopic mode using the instrument Dolores (scale = 0.275
\AA pxl$^{-1}$, dispersion = 0.8 \AA pxl$^{-1}$, $\lambda \in (6200,7800)$
\AA). In Table 1 we list the properties of some of the observed
galaxies.

\begin{table}
\centering
\scriptsize{\caption{Properties of the observed galaxies}
\vskip -0.3truecm
\begin{tabular}{l c c c c c}
      &  &  &  &  & \\
Name & Type & D (Mpc) & M (mag) & $\mathrm{\mu}_{o}$ (mag arcsec$^{-1}$) & h
(kpc)\\
\tableline
UGC 4325    & SA(s)m & 10.1 & -18.1(R) & 21.6(R) & 1.7(R)\\
UGC 4499    & SABdm  & 13.0 & -17.8(R) & 21.5(R) & 2.0(R)\\
UGC 11861   & SABdm  & 25.1 & -20.8(R) & 21.4(R) & 6.1(R)\\
LSBC F571-8 & Sc     & 48.0 & -17.6(B) & 23.9(B) & 5.2(B)\\
\end{tabular}}
\end{table}

\begin{figure}[!h]
\centering
\vskip -0.3truecm
\small{\plottwo{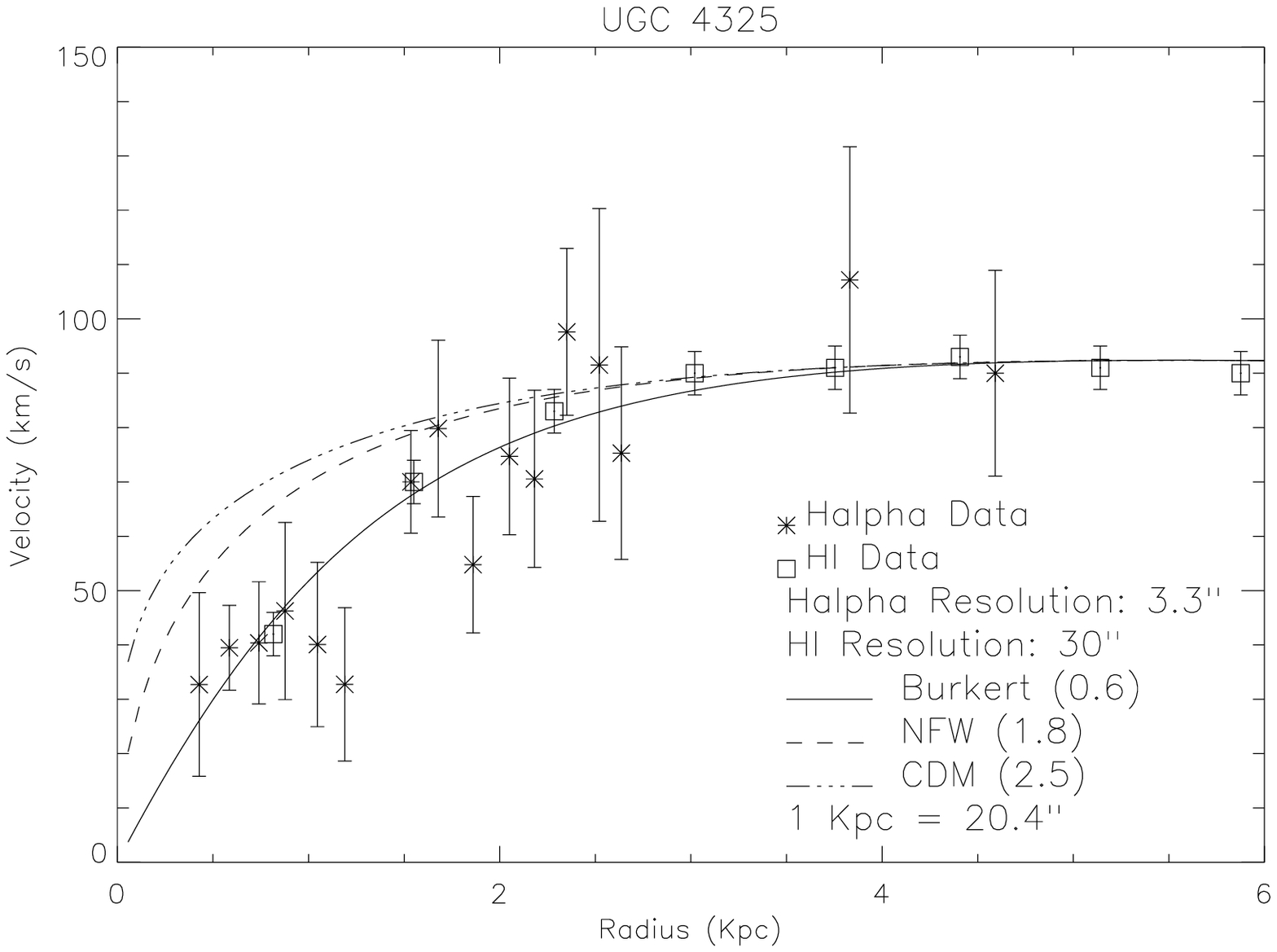}{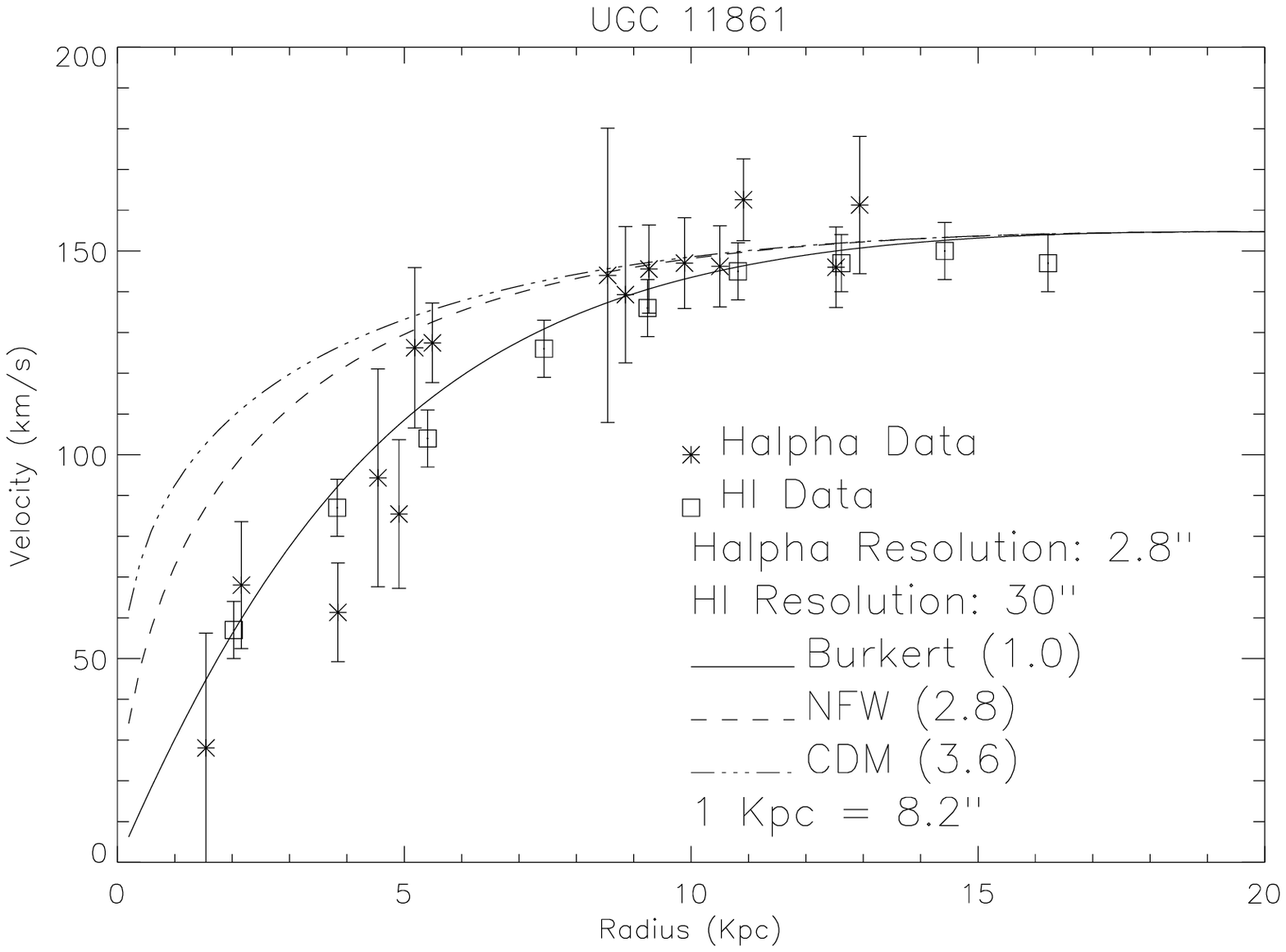}}
\vskip -0.4truecm
\end{figure}

\begin{figure}[!h]
\centering
\small{\plottwo{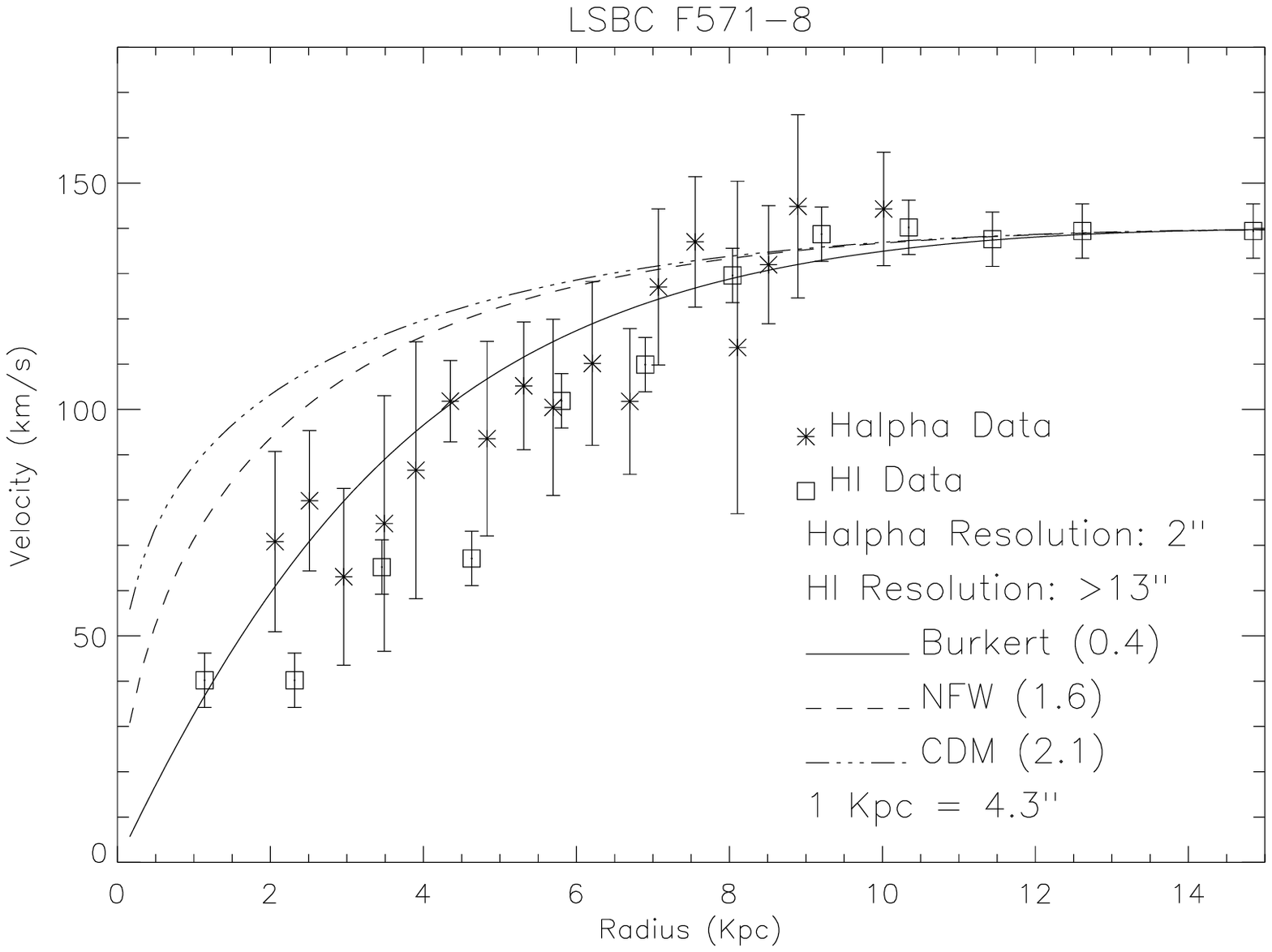}{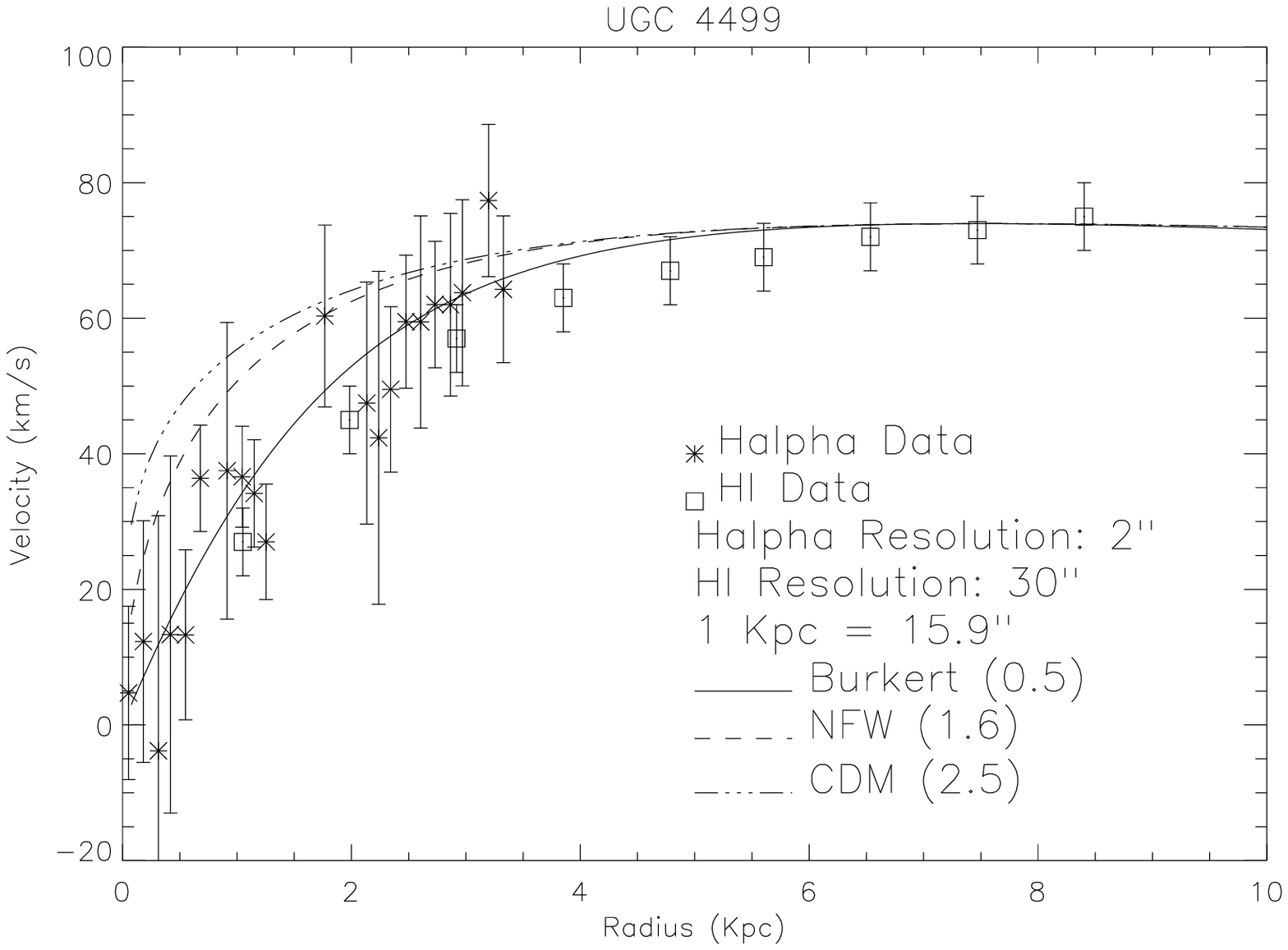}}
\vskip -0.7truecm
\end{figure}

Our H$\alpha$ data
(stars in the figures) are plotted together with HI data (from literature, empty squares). As one can see,
the HI data suffer of very little beam smearing. The solid lines are the
best fits obtained by using the Burkert profile; the dashed lines are NFW
profiles and the dotted-dashed lines are Moore profiles (CDM model). The
numbers in parenthesis represent the normalized $\chi^{2}$ for each fit. For
each galaxy is stated the spatial resolution for both the H$\alpha$ and the HI 
data. For all the four galaxies, the Burkert profile is the best fit, while
the Moore profile does not match data. New observations with high resolution
new technology grating (VPHG, Conconi et al. 2001) are being planned.

\section{Conclusions}
Our preliminary analysis of H$\alpha$ observations seems to rule out the CDM
model in the inner regions of LSB and dwarf galaxies, showing evidence of soft 
cores.

\end{document}